\documentclass[reprint,pre,aps,superscriptaddress,floatfix]{revtex4-2}
\usepackage{graphicx,amsmath,amssymb}
\usepackage[usenames]{color}
\usepackage{cancel}
\usepackage{ulem}
\usepackage{hyperref}
\hypersetup{colorlinks,
 linkcolor=blue,%
 citecolor=blue,%
 urlcolor=blue
}

\begin{document}

\title{Synchronizability in randomized weighted  simplicial complexes}

\author{S. Nirmala Jenifer}
\affiliation{Department of Physics, Bharathidasan University, Tiruchirappalli 620024, Tamil Nadu, India}

\author{Dibakar Ghosh}\email{dibakar@isical.ac.in}
\affiliation{Physics and Applied Mathematics Unit, Indian Statistical Institute, 203 B. T. Road, Kolkata - 700108, India}

\author{Paulsamy Muruganandam}\email{anand@bdu.ac.in}
\affiliation{Department of Physics, Bharathidasan University, Tiruchirappalli 620024, Tamil Nadu, India}

\date{\today}

\begin{abstract}
We present a formula for determining synchronizability in large, randomized and weighted simplicial complexes. This formula leverages eigenratios and costs to assess complete synchronizability under diverse network topologies and intensity distributions. We systematically vary coupling strengths (pairwise and three-body), degree and intensity distributions to identify the synchronizability of these simplicial complexes of the identical oscillators with natural coupling. We focus on randomized weighted connections with diffusive couplings and check synchronizability for different cases. For all these scenarios, eigenratios and costs reliably gauge synchronizability, eliminating the need for explicit connectivity matrices and eigenvalue calculations. This efficient approach offers a general formula for manipulating synchronizability in diffusively coupled identical systems with higher-order interactions simply by manipulating degrees, weights, and coupling strengths. We validate our findings with simplicial complexes of R\"ossler oscillators and confirm that the results are independent of the number of oscillators, connectivity components and distributions of degrees and intensities. Finally, we validate the theory by considering a real-world connection topology using  chaotic R\"ossler oscillators.
\end{abstract}

\maketitle

\section{Introduction}
Complex systems arise when a large number of dynamical units interact with each other, giving rise to emergent properties distinct from those of the individual subsystems \cite{Boccaletti2006}. Such systems are ubiquitous, found naturally in entities like the brain and man-made ones like the internet or financial markets \cite{Kwapien2012}. The study and modelling of these systems have been a longstanding focus of researchers~\cite{Kuramoto1975, Iacopini2019, Strogatz1993}. Complex networks represent the dynamical units, and their interactions as nodes and links are a common way to model such systems. However, these networks typically only account for pairwise interactions, whereas many complex systems involve interactions among three or more units. To gain a more comprehensive understanding of these systems, one must incorporate higher-order networks, such as hypergraphs and simplicial complexes, to capture these higher-order interactions \cite{Majhi2022}.

Simplicial complexes are used to model complex systems with higher-order interactions \cite{Iacopini2019, Battiston2021, Battiston2020, Boccaletti2023}. A simplicial complex consists of $d$-simplices, where $d$ is the dimension of the simplex. A $0$-dimensional simplex is a node, a $1$-dimensional simplex is a link, a $2$-dimensional simplex is a triangle, and so on. A $d$-dimensional simplicial complex consists of simplices up to dimension $d$. They provide a good representation of complex systems with higher-order interactions \cite{Grilli2017, Mayfield2017, Battiston2021, Alvarez2021}. In most studies, these simplicial complexes are unweighted, meaning all the links and triangles have the same weight \cite{Anwar2022, Anwar2023}. However, it has been observed that in most real-world systems, this is not feasible. For instance, in collaboration networks, if there is a three-author paper, not every pair of authors has to produce research papers by themselves. But in unweighted simplicial complexes, if there is a three-author paper, then there is also the presence of all the possible two-author research papers, and this leads to information loss, which one could avoid by introducing proper weights for links \cite{Baccini2022}. Courtney and Bianconi \cite{Courtney2017} have developed a model to distribute weights to links in terms of bare affinity weights and topological weights. Every simplex has bare-affinity weights, while the topological weights represent whether a particular link has a contribution other than being part of a triangle. Therefore, weighted simplicial complexes provide a more realistic representation of higher-order networks.

After accurately modelling complex systems, we can better understand their emergent properties, such as synchronization. As a result, we can control the synchronization that occurs in natural or artificial systems. Synchronization occurs when individual dynamical systems adjust their properties to have common dynamics \cite{Kuramoto1975}. Neuronal synchronization can cause epilepsy, while synchronization from a healthy to an infected state can cause an epidemic outbreak, and so on. Synchronization can be both constructive and destructive. However, it is possible to promote synchronization in desired systems and inhibit it where it leads to destruction. This can be achieved by determining under what circumstances the system goes into synchronization and desynchronization and whether synchronization is possible. Two main parameters are used to calculate the synchronizability of the system: eigenratio and the cost of the connections of nodes in the system. The lower the eigenratio and cost, the system has more synchronizability \cite{Zhou2006}.

To calculate the eigenratio and costs, we need to construct the connectivity matrices, such as adjacency matrices and Laplacian matrices, and then find the eigenvalues of these matrices. Writing connectivity matrices for large, complex systems, such as connection networks, collaboration networks, and neuronal networks, can be challenging, and collecting the necessary data is not an easy task. It becomes even more difficult when we include higher-order interactions, as we need to construct the adjacency tensors for these interactions. Even for three-body interactions, the adjacency tensor contains $N \times N \times N$ elements. In 2006, Zhou {\it et al.} \cite{Zhou2006} developed an approach to calculate eigenratios and costs based on the heterogeneity of the system intensities. This approach significantly reduced the computational cost of constructing connectivity matrices and evaluating their eigenvalues. We extend this approach to randomized weighted simplicial complexes. As a result, we can determine the synchronizability parameters solely from the heterogeneity of the intensities, specifically the maximum and minimum intensities of a node as part of links ($d=1$), triangles ($d=2$), and so on and from their coupling strengths. In this paper, we discuss the mathematical formulation, derive general formulas for determining the synchronization without explicitly computing the eigenvalues of huge matrices and corroborate with numerical results. We validate the results by applying the general formula to  R\"ossler oscillators and real-world connection networks. We check the emergence of complete synchronization in diffusively coupled identical R\"ossler oscillators, and we show that the results are independent of the number of oscillators in the networks, the components through which they are connected and also the distributions of the degrees and intensities.

\section{Analytical Results} 
In order to derive the general formulas for the eigenratio and the cost, we first write the dynamical equations for the weighted simplicial complex. Then, we deduce the corresponding variational equations and modify these equations in terms of the effective matrix $M$. The eigenvalues of the effective matrix determine the stability of the synchronized state. Since the effective matrix is a zero row sum matrix, the first eigenvalue will be zero, corresponding to the mode along the synchronization manifold. We can find the synchronizability of the system from the ratio between $\lambda_N$ and $\lambda_2$. For simplicity, we consider the simplicial complex of dimension $2$, which we can extend to any dimension.

We consider a simplicial complex of identically coupled oscillators, described by the following equations,
\begin{align}
\mathbf{ \dot{x}}_i = & \mathbf{f}(\mathbf{x}_i) + \sigma_1 \sum_{j=1}^N a_{ij}^{(1)} \omega_{ij}^{(1)} \mathbf{g}^{(1)}(\mathbf{x}_i,\mathbf{x}_j) \notag \\ & + \sigma_2{\sum_{j=1}^N}{\sum_{k=1}^N} a_{ijk}^{(2)} \omega_{ijk}^{(2)} \mathbf{g}^{(2)}(\mathbf{x}_i,\mathbf{x}_j,\mathbf{x}_k), \label{eq:1}
\end{align}
where $\mathbf{x}_i$ is the state vector of $i$-th oscillator of dimension $m$, and $\mathbf{f}(\mathbf{x}_i)$ represents the dynamics of the uncoupled oscillators. ${\omega_{ij}}^{(1)}$ and $\omega_{ijk}^{(2)}$ are the topological weights of links and triangles, respectively. $a_{ij}^{(1)}$ and $a_{ijk}^{(2)}$ are the elements of the adjacency matrix $A^{(1)}$ and adjacency tensor $A^{(2)}$. $a_{ij}^{(1)} = 1$ if the nodes $i$ and $j$ form a link, or $a_{ij}^{(1)} = 0$ otherwise. Also $a_{ijk}^{(2)} = 1$ if the nodes $i$, $j$, and $k$ form a triangle, or $a_{ijk}^{(2)} = 0$ otherwise. Here $\sigma_1$ and $\sigma_2$ are the coupling strengths of the links (pairwise) and triangles (non-pairwise), respectively. Here, $\mathbf{g}^{(1)}(\mathbf{x}_i,\mathbf{x}_j)$ and $\mathbf{g}^{(2)}(\mathbf{x}_i,\mathbf{x}_j,\mathbf{x}_k)$ are the synchronization noninvasive functions \cite{Battiston2021} and are chosen in the forms $\mathbf{g}^{(1)}(\mathbf{x}_i,\mathbf{x}_j)=\mathbf{h}^{(1)} (\mathbf{x}_j) - \mathbf{h}^{(1)}(\mathbf{x}_i)$ and $\mathbf{g}^{(2)}(\mathbf{x}_i,\mathbf{x}_j,\mathbf{x}_k)=\mathbf{h}^{(2)}(\mathbf{x}_j, \mathbf{x}_k) - \mathbf{h}^{(2)}(\mathbf{x}_i, \mathbf{x}_i)$. At the synchronization state, they tend to zero, making the dynamics of the system resemble that of the uncoupled oscillators.

For the above diffusive coupling functions, we can rewrite the above equation as
\begin{align}
\mathbf{ \dot{x}}_i = & \, \mathbf{f}(\mathbf{x}_i) + \sigma_1\sum_{j=1}^N a_{ij}^{(1)} \omega_{ij}^{(1)} \left[\mathbf{h}^{(1)} (\mathbf{x}_j) - \mathbf{h}^{(1)}(\mathbf{x}_i) \right] \notag \\ & \, + \sigma_2{\sum_{j=1}^N}{\sum_{k=1}^N}a_{ijk}^{(2)} \omega_{ijk}^{(2)} \left[\mathbf{h}^{(2)}(\mathbf{x}_j, \mathbf{x}_k) - \mathbf{h}^{(2)}(\mathbf{x}_i, \mathbf{x}_i) \right], \label{eq:2:0}
\end{align}
where $\mathbf{h}^{(1)}(\mathbf{x}_j)$ and $\mathbf{h}^{(2)}(\mathbf{x}_j, \mathbf{x}_k) $ are coupling functions that couples nodes in links and triangles, respectively. The choice of coupling functions affects the synchronizability of the system \cite{Battiston2021}. So we have to choose the coupling functions in a way that the oscillators tend to synchronize and we can write,
\begin{align}
S_i^{(1)} = \sum_{j=1}^N{a_{ij}}^{(1)} \omega_{ij}^{(1)}\, \;\; \mbox{and}\;\; 
S_i^{(2)}=\frac{1}{2}\sum_{j=1}^N\sum_{k=1}^N a_{ijk}^{(2)}\omega_{ijk}^{(2)}. \notag
\end{align}
Here $S_i^{(1)}$ and $S_i^{(2)} $ are the intensities of node $i$ to form links and triangles, i.e., the number of weighted links and weighted triangles incident on node $i$. For randomized simplicial complexes with $K^{(1)}_{\mbox{min}} \gg 1$, we can use mean-field approximation and this equation can be rewritten as,
\begin{align}
\dot{\mathbf x}_i = & \, \mathbf{f}(\mathbf{x}_i) + 
\sigma_1\frac{S_i^{(1)}}{K_i^{(1)}}\sum_{j=1}^N a_{ij}^{(1)}\left(\mathbf{h}^{(1)}(\mathbf{x}_j) - \mathbf{h}^{(1)}(\mathbf{x}_i)\right) 
\notag \\ & + \sigma_2\frac{S_i^{(2)}}{K_i^{(2)}} 
 {\sum_{j=1}^N}{\sum_{k=1}^N}a_{ijk}^{(2)} \left(\mathbf{h}^{(2)}(\mathbf{x}_j,\mathbf{x}_k) - \mathbf{h}^{(2)}(\mathbf{x}_i,\mathbf{x}_i) \right). \label{eq:intensity} 
\end{align}
Here, $K_i^{(1)}$ is the degree of pairwise interactions, i.e., the total number of links that a node participates in. $K_i^{(2)}$ is the total number of triangles that a node participates in, also called the degree for three-body interactions. Now, we consider pairwise and non-pairwise local mean fields as
\begin{align}
\mathbf{\overline{H}}^{(1)}{(\mathbf{x}_i)} = \frac{1}{K_i^{(1)}}\sum_{j=1}^N a_{ij}^{(1)} \mathbf{h}^{(1)}(\mathbf{x}_j) \notag
\end{align}
due to the interaction between the nodes through links and 
\begin{align}
\mathbf{\overline{H}}^{(2)}{(\mathbf{x}_i,\mathbf{x}_i})= \frac{1}{K_i^{(2)}}{\sum_{j=1}^N}{\sum_{k=1}^N} a_{ijk}^{(2)} \mathbf{h}^{(2)}(\mathbf{x}_j,\mathbf{x}_k) \notag
\end{align}
due to the interaction between nodes through triangles. Then, Eq.~\eqref{eq:intensity} can be written in terms of mean field as,
\begin{align} 
\mathbf{\dot{x}}_i = &\, \mathbf{f}(\mathbf{x}_i) + \sigma_1S_i^{(1)}\left(\mathbf{\overline{H}}^{(1)}(\mathbf{x_i})-\mathbf{h^{(1)}}(\mathbf{x}_i)\right) \notag \\ & + \sigma_2{S_i^{(2)}}\left( \mathbf{\overline{H}}^{(2)}{(\mathbf{x}_i,\mathbf{x}_i}) - \mathbf{h}^{(2)}(\mathbf{x}_i,\mathbf{x}_i) \right). \label{eq:2}
\end{align}
So, the interaction between nodes in a large randomized weighted simplicial complex can be approximated as the interaction of a node with a mean field. This mean field is not only generated by the interaction of links, as in the case of complex networks, but also by the interaction of nodes as triangles.

According to the condition for natural coupling, close to the synchronized state, the interaction between nodes in links and triangles will be similar since, upon synchronization, the two nodes that are in the same state can be considered as a single node, and the three-body interactions reduce to the two-body interactions \cite{Battiston2021}.
The above assumptions enable us to write $\mathbf{h}^{(2)}(\mathbf{x}, \mathbf{x}) = \mathbf{h}^{(1)}(\mathbf{x})$ and $\overline{\mathbf H}^{(2)}{(\mathbf{x}, \mathbf{x})} = \overline{\mathbf H}^{(1)}{(\mathbf{x})} = \mathbf{H(x)}$.
Equation \eqref{eq:2} can be rewritten as,
\begin{align} \label{eq:3}
\mathbf{\dot{x}}_i = \mathbf{f}(\mathbf{x}_i) + \left(\sigma_1S_i^{(1)} + \sigma_2S_i^{(2)}\right) \left(\mathbf{H}(\mathbf{x})-\mathbf{h}^{(1)}(\mathbf{x}_i) \right).
\end{align}
The variational equation of the above Eq.~\eqref{eq:3} is of the form
\begin{align} 
\delta\mathbf{ \dot{x}}_i = & \left[ J\mathbf{f}(\mathbf{x}^s) - \left(\sigma_1S_i^{(1)} + \sigma_2S_i^{(2)} \right)J\left(\mathbf{h}^{(1)}(\mathbf{x}^s)\right) \right] \delta\mathbf{x}_i, \label{variational} 
\end{align}
where $\mathbf{x}^s$ is the synchronization state.
The eigenvalues are approximately equal to $\sigma_1S_i^{(1)} + \sigma_2S_i^{(2)}$, $i = 1, 2, \ldots, N$.
From the above equation, we can write the eigenratio \cite{Zhou2006} as,
\begin{align}
R \approx \frac{\sigma_1 S_{\mbox{max}}^{(1)} + \sigma_2 S_{\mbox{max}}^{(2)}}{\sigma_1 S_{\mbox{min}}^{(1)} + \sigma_2 S_{\mbox{min}}^{(2)}},
\end{align}
where $\sigma_1 S_{\mbox{max}}^{(1)}+\sigma_2 S_{\mbox{max}}^{(2)} \ge \mbox{max}\{ \sigma_1 S_1^{(1)}+\sigma_2 S_1^{(2)}, \sigma_1 S_2^{(1)}+\sigma_2 S_2^{(2)}, \cdots, \sigma_1 S_N^{(1)}+\sigma_2 S_N^{(2)}\}$ will give the maximum bound for eigenvalues and similar for minimum $\sigma_1 S_{\mbox{min}}^{(1)}+\sigma_2 S_{\mbox{min}}^{(2)} \le \mbox{min}\{ \sigma_1 S_1^{(1)}+\sigma_2 S_1^{(2)}, \sigma_1 S_2^{(1)}+\sigma_2 S_2^{(2)}, \cdots, \sigma_1 S_N^{(1)}+\sigma_2 S_N^{(2)}\}$. These maximum and minimum bounds may not necessarily belong to any particular node index $i$. It further simplifies the problem as it is sufficient to consider only the maximum and minimum values of the intensities to calculate the eigenratio and it eliminates the necessity to calculate all the eigenvalues. It would be of great advantage when $N$ is very large.
 
For the case of pairwise interaction only ($\sigma_2=0.0$), the eigenratio $R$ is very similar to the previous result \cite{Zhou2006}. For non-zero pairwise coupling strength ($\sigma_1 \neq 0.0,$), we can rewrite the above equation as,
\begin{align} \label{eq:5}
R \approx \frac{S_{\mbox{max}}^{(1)} + \frac{\sigma_2}{\sigma_1}S_{\mbox{max}}^{(2)}}{S_{\mbox{min}}^{(1)} + \frac{\sigma_2}{\sigma_1}S_{\mbox{min}}^{(2)}}.
\end{align}
In the similar way, we can derive the eigenratio for the $d$-dimensional simplical complex as,
\begin{align}\label{eq:5.1}
R \approx \frac{S_{\mbox{max}}^{(1)} + \frac{\sigma_2}{\sigma_1}S_{\mbox{max}}^{(2)}+ \cdots + \frac{\sigma_d}{\sigma_1}S_{\mbox{max}}^{(d)}}{S_{\mbox{min}}^{(1)} + \frac{\sigma_2}{\sigma_1}S_{\mbox{min}}^{(2)} + \cdots +\frac{\sigma_d}{\sigma_1}S_{\mbox{min}}^{(d)}}.
\end{align} %
Thus, the eigenratio of a randomized simplicial complex depends on the coupling strengths (pairwise and non-pairwise), and maximum and minimum intensities of the nodes. If we assume all the non-pairwise coupling strengths are identical to the pairwise coupling strength (i.e., $\sigma_2=\sigma_3=\cdots=\sigma_d=\sigma_1$), then the eigenratio $R$ is independent on the coupling strengths and depends on the maximum and minimum intensities as observed for pairwise weighted random networks \cite{Zhou2006}. 

To find the tight bounds of the above mean-field approximation, we can write Eq.~\eqref{eq:3} as,
\begin{align} \label{eq:6}
\mathbf{ \dot{x}_i} = & \, \mathbf{F}(\mathbf{x}_i) - \sigma_1\sum_{j=1}^N G_{ij}^{(1)}\mathbf{H}^{(1)}(\mathbf{x}_j) 
- \sigma_2\sum_{j=1}^N G_{ij}^{(2)}\mathbf{H}^{(2)}(\mathbf{x}_j),
\end{align}
where $G_{ij}^{(1)}$ and $G_{ij}^{(2)}$ are the elements of matrices $G^{(1)}= S^{(1)}D^{(1)-1}L^{(1)}$ and $G^{(2)}=S^{(2)}D^{(2)-1}L^{(2)}$, respectively. 
Here $L^{(1)}$ and $L^{(2)}$ are generalized Laplacian matrices, while $S$ and $D$ are diagonal matrices of strengths and generalized degrees, respectively. It is a good point that we have separated weights from topology.
We can write normalized Laplacian matrices as $\overline{L}^{(1)} = D^{-1}L^{(1)}$ and $\overline{L}^{(2)} = D^{-1}L^{(2)}$, then $G^{(1)} = S^{(1)}\overline{L}^{(1)}$ and $G^{(2)} = S^{(2)}\overline{L}^{(2)}$. Thus, the largest and smallest eigenvalues of the matrices $G^{(1)}$ and $G^{(2)}$ are bounded by the eigenvalues of $\overline{L}^{(1)}$ and $\overline{L}^{(2)}$. We can write the linearized variational Eq. \eqref{eq:6} by following the procedure outlined in \cite{Gambuzza2021} and is given by
\begin{align} 
\delta\dot{\mathbf x}_i = \left(J\mathbf{f}(\mathbf{x}^s) - \sum_{j=1}^N\left[\sigma_1 G_{ij}^{(1)} + \sigma_2 G_{ij}^{(2)}\right]J\left(\mathbf{h}^{(1)}(\mathbf{x}^s) \right)\right)\delta{\mathbf x}_i. \label{eq:7}
\end{align}
We can define an effective matrix $M$ as,
\begin{align} \label{eq:8}
 M_{ij}= G^{(1)}_{ij} + \frac{\sigma_2}{\sigma_1}G^{(2)}_{ij}.
\end{align}
Equation~\eqref{eq:7} can be rewritten, in terms of effective matrix $M$, as 
\begin{align} \label{eq:9}
 \delta\mathbf{ \dot{x}}_i = \left[J\mathbf{f}\left(\mathbf{x}^s\right) - \sigma_1\sum_{j=1}^N M_{ij} J\left(\mathbf{h}^{(1)}\mathbf{x}^s\right)\right]\delta\mathbf{x}_i.
\end{align}
The eigenvalues of the matrix $M$ depend on the ratio of coupling strengths, degrees and intensities, and the eigenratio of this matrix can be expressed as $R=\lambda_N / \lambda_2,$ where $\lambda_2$ and $\lambda_N$ are the 2nd and $N$-th eigenvalues of $M$ such that $0=\lambda_1\le \lambda_2 \le \lambda_3\le \cdots \le \lambda_N$.

Another measure of synchronizability is the cost $C$ involved in the coupling of nodes in a simplicial complex, and it is the total strength of connections of all nodes. It can be written as,
\begin{align}
C = \sigma_1\sum_{i=1}^N{S_{i}}^{(1)} + \sigma_2\sum_{i=1}^NS_{i}^{(2)},
\end{align}
By using the eigenvalues of $M$ in Eq.~\eqref{eq:9}, the normalized cost can be written as,
\begin{align} \label{eq:12}
 C_0 = \frac{C}{N\alpha_1} = \frac{\Omega}{\lambda_2},
\end{align}
where $\alpha_1 = \sigma_1\lambda_2(M) $ and mean intensity
\begin{align} \label{eq:12:1}
 \Omega = \frac{1}{N}\left( \sum_{i=1}^N S_{i}^{(1)} + \frac{\sigma_2}{\sigma_1}\sum_{i=1}^NS_{i}^{(2)} \right).
\end{align}
From Eq.~\eqref{variational}, the normalized cost will be,
\begin{align} \label{eq:12:2}
C_0 \approx \frac{\Omega}{{S_{\mbox{min}}^{(1)} + r S_{\mbox{min}}^{(2)}}}
\end{align}
where $r={\sigma_2}/{\sigma_1}$.
We can verify Eq.~\eqref{eq:5} and Eq.~\eqref{eq:12:2} numerically by computing the eigenvalues of the effective matrix $M$, and we can obtain an explicit bound from the eigenvalues of $\overline{L}^{(1)}$ and $\overline{L}^{(2)}$. Using this bound, we can derive formulas for $R$ and $C_0$ as a function of $S_{\mbox{max}}^{(1)}$, $S_{\mbox{min}}^{(1)}$ and the generalized degrees.

We can demonstrate that the upper and lower bounds of the nonzero eigenvalues of the effective matrix $M$ are given by the eigenvalues $\mu^{(1)}_l$ and $\mu^{(2)}_l$ of the matrices $G^{(1)}$ and $G^{(2)}$, respectively, as follows,
\begin{align} 
& {S_{\mbox{min}}^{(1)}\mu_2^{(1)} + rS_{\mbox{min}}^{(2)}\mu_2^{(2)}} \le \lambda_2 \le {S_{\mbox{min}}^{(1)} + r S_{\mbox{min}}^{(2)}},
 \label{eq:13} \\
& {S_{\mbox{max}}^{(1)} + r S_{\mbox{max}}^{(2)}} \le \lambda_N \le {S_{\mbox{max}}^{(1)}\mu_N^{(1)} + rS_{\mbox{max}}^{(2)}\mu_N^{(2)}}.\label{eq:14}
\end{align}
For sufficiently random simplicial complexes, the spectra of the matrices $G^{(1)}$ and $G^{(2)}$ tend to follow the semicircle law \cite{Zhou2006} and we can write $\mu_2^{(1)} \approx 1 - 2/\sqrt{K^{(1)}}$, $\mu_N^{(1)} \approx 1 + 2/\sqrt{K^{(1)}}$, $\mu_2^{(2)} \approx 1 - 2/\sqrt{K^{(2)}}$ and $\mu_N^{(2)} \approx 1 + 2/\sqrt{K^{(2)}}$, 
provided that $K_{\mbox{min}}^{(d)} \gg \sqrt{K^{(d)}} $. Here $K^{(d)}$ is the mean degree of the $d$-simplex. By introducing these in Eq. \eqref{eq:13} and Eq. \eqref{eq:14}, we get the following approximations for the bounds of $R$ and $C_0$ as
\begin{align} \label{eq:15}
 \frac{S_{\mbox{max}}^{(1)} +r S_{\mbox{max}}^{(2)}}{S_{\mbox{min}}^{(1)} + rS_{\mbox{min}}^{(2)}} & \le R \notag \\ & \le \frac{1 + \frac{2}{\sqrt{K^{(1)}}}}{1 - \frac{2}{\sqrt{K^{(1)}}}} \left( \frac{S_{\mbox{max}}^{(1)} +r\frac{1 + \frac{2}{\sqrt{K^{(2)}}}}{1 + \frac{2}{\sqrt{K^{(1)}}}}S_{\mbox{max}}^{(2)}}{S_{\mbox{min}}^{(1)} + r\frac{1 - \frac{2}{\sqrt{K^{(2)}}}}{1 - \frac{2}{\sqrt{K^{(1)}}}}S_{\mbox{min}}^{(2)}} \right).
\end{align}
\begin{align} \label{eq:15:2} 
 \frac{\Omega}{S_{\mbox{min}}^{(1)} + r S_{\mbox{min}}^{(2)}} & \le C_0 \notag \\ & \le \frac{1}{1 - \frac{2}{\sqrt{K^{(1)}}}} \left( \frac{\Omega}{S_{\mbox{min}}^{(1)} + r\frac{1 - \frac{2}{\sqrt{K^{(2)}}}}{1 - \frac{2}{\sqrt{K^{(1)}}}}S_{\mbox{min}}^{(2)}} \right).
\end{align}
For a given value of $K$, the synchronizability of randomly weighted simplicial complexes with a large value of $K_{\mbox{min}}^{(1)}$ is expected to follow a general formula as follows,
\begin{align} \label{eq:15:3}
R = A_R\left(\frac{S_{\mbox{max}}^{(1)} + B_{R_1} S_{\mbox{max}}^{(2)}}{S_{\mbox{min}}^{(1)} + B_{R_2}S_{\mbox{min}}^{(2)}} \right),
\end{align}
and 
\begin{align} \label{eq:15:4}
C_0 = A_C\left(\frac{\Omega}{S_{\mbox{min}}^{(1)} + B_{R_2}S_{\mbox{min}}^{(2)}} \right),
\end{align}
where $A_R = \frac{1 + 2/\sqrt{K^{(1)}}}{1 - 2/\sqrt{K^{(1)}}}$ and $ A_C =\frac{1}{1 - 2/\sqrt{K^{(1)}}}$. Here, $A_R \rightarrow 1$ and $A_C \rightarrow 1$ in the limit $K^{(1)} \rightarrow \infty $. Also, $B_{R_1} = r\frac{1 + 2/\sqrt{K^{(2)}}}{1 + 2/\sqrt{K^{(1)}}} $ and $ B_{R_2} = r\frac{1 - 2/\sqrt{K^{(2)}}}{1 - 2/\sqrt{K^{(1)}}} $. $B_{R_1} \rightarrow r$ and $ B_{R_2} \rightarrow r$ in the limit $K^{(d)} \rightarrow \infty (d=1,2)$. They are well agreed with Eqs. \eqref{eq:5} and \eqref{eq:12:2}. 

For $N$-dimensional simplicial complex, 
\begin{align} \label{eq:15:5}
R = A_R\left(\frac{S_{\mbox{max}}^{(1)} +\sum_{d=2}^N B_{R_1}^{(d)} S_{\mbox{max}}^{(N)}}{S_{\mbox{min}}^{(1)} + \sum_{d=2}^N B_{R_2}^{(d)} S_{\mbox{min}}^{(N)}} \right)
\end{align}
and 
\begin{align} \label{eq:15:6}
C_0 = A_C\left(\frac{\Omega}{S_{\mbox{min}}^{(1)} + \sum_{d=2}^N B_C^{(d)} S_{\mbox{min}}^{(N)}} \right),
\end{align}
where 
\begin{align}
B_{R_1}^{(d)} = r\frac{1 + 2/\sqrt{K^{(d)}}}{1 + 2/\sqrt{K^{(1)}}},\;\; B_{R_2}^{(d)
} = r\frac{1 - 2/\sqrt{K^{(d)}}}{1 - 2/\sqrt{K^{(1)}}}.
\end{align}

\section{Numerical Results}

\subsection{Verification of the general formula}
We compare the expression of $R$ as a function of eigenvalue ratio ${\lambda_N}/{\lambda_2}$ and $C_0$ as a function of the ratio $\Omega/\lambda_2$ and found that they are almost identical when $K^{(1)}_{\mbox{max}} \gg 1$.

To conduct our analysis, the parameter values are set as follows: $N = 10^3$, $\sigma_1 = 0.001$, $\sigma_2 = 0.01$. We arbitrarily fix the values of the minimum and maximum intensities of the node for pairwise $(S^{(1)}_{\mbox{min}}, S^{(1)}_{\mbox{max}})$ and non-pairwise $(S^{(2)}_{\mbox{min}}, S^{(2)}_{\mbox{max}})$ interactions. In each iteration of the simulation, we add $100$ to $S^{(1)}_{\mbox{max}}$ and keep all other values constant. The simulation is run $10^3$ times. We first consider a simplicial complex with a node $i$ connected to all other nodes, resulting in $K^{(1)}_{\mbox{max}} = N - 1$. The maximum number of triangles incident on a node is $(N-1) \times (N-2)$, and the maximum number of triangles incident on a link is $(N-2)$. Using this structure, we generate the Laplacian matrices $L^{(1)}$ and $L^{(2)}$. Finally, we uniformly distribute $S$ among the nodes. All the triangles are considered as nodes having three-body interactions. 

We calculate the eigenratio $R$ from the effective matrix $M$ using Eq.~\eqref{eq:8} and compare it with Eq.~\eqref{eq:15:3}. We then plot the values of $R$ against the corresponding values of the eigenratio in terms of $S$ calculated from the general formula, which reveals a linear relationship between $R$ and the eigenratio as a function of $S$. %
\begin{figure}[!ht]
\centering\includegraphics[width=0.99\linewidth]{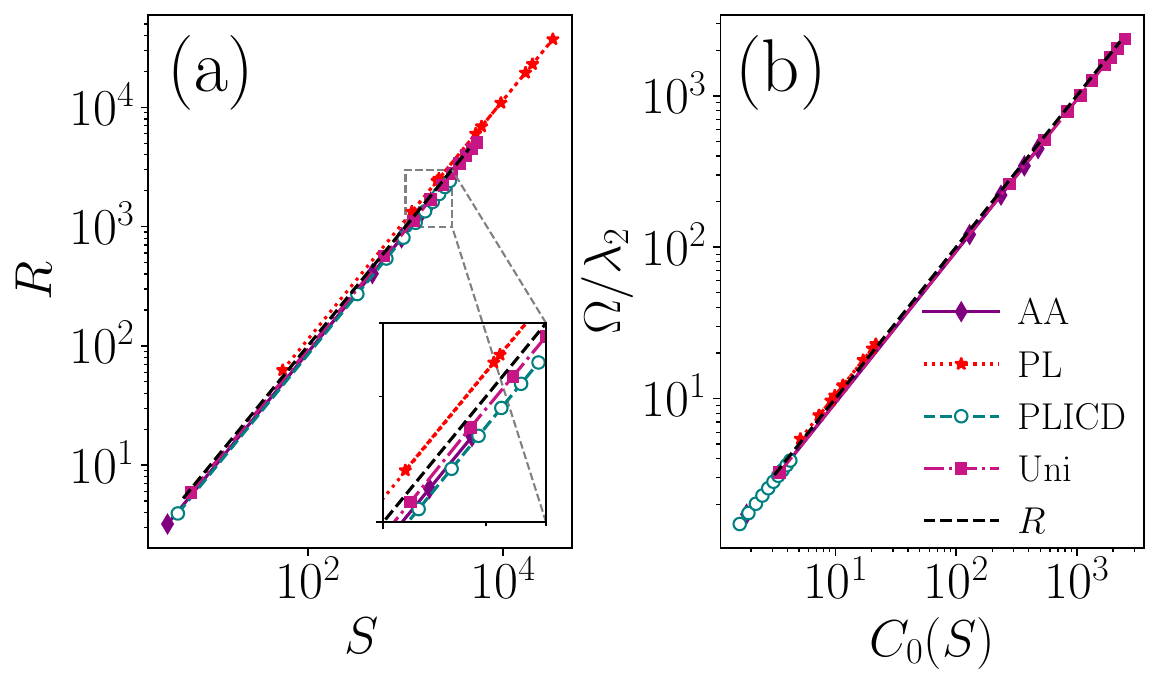} 
\caption{(a) $R$ (calculated using the eigenvalues of matrix $M$) vs $S$ (eigenratio calculated using the general formula), and (b) $\Omega/\lambda_2$ vs $C_0$, expressed in terms of intensities, as computed using the general formula for different coupling schemes and distributions: (AA) all-to-all coupling, (PL) power law distribution, (PLICD) power law distribution with intensities correlated to degrees, (Uni) uniform distribution, and ($R$) $R$ computed using the eigenvalues of $M$.}
\label{fig:1}
\end{figure}
Similarly, we calculate the normalized cost $C_0$ using Eq. \eqref{eq:8} and compare it with Eq. ~\eqref{eq:15:4}. We then plot the values of $\Omega/\lambda_2$ against the corresponding values of $C_0$, which reveals a linear relationship between them, which is shown dashed-purple line with diamond symbol (AA) in Fig.~\ref{fig:1}. 

One may note that the most realistic networks are scale-free networks, where the degrees are distributed according to a power law \cite{Barabasi2003, Adamic2000}. This power-law distribution means that only a small portion of nodes has very high degrees, while a majority of nodes have relatively small degrees. Therefore, we construct a simplicial complex with power-law distributed degrees and intensities; however, they are not related. We achieve this by distributing degrees and intensities to nodes using a power-law distribution. In Fig. \ref{fig:1}, we also show these results by the dotted-red line with stars (PL) and the green-dashed line with empty circles (PLICD).

Next, we consider the scale-free distribution with intensities correlated to degrees. We construct this by distributing the intensities, which are proportional to degrees. Hence, the node with the maximum degree will receive the maximum intensity, and the node with the minimum degree will receive the minimum intensity. We apply this procedure to both three-body interactions and two-body interactions. Subsequently, we plot the graphs for $R$ and $C_0$ corresponding to this distribution.
The results demonstrate a linear relationship between the values of the eigenratio and cost calculated from the general formula. This linear relationship also holds for the uniform distribution, as depicted by the dash-dotted pink line with square symbols (Uni) in Fig.~\ref{fig:1}.
From Fig.~\ref{fig:1}, we can conclude that the general formula holds true regardless of the distribution of degrees and intensities. Further, we observe that the values obtained from the general formula are nearly identical to that obtained by finding the eigenvalues of the effective matrix $M$ (blue dashed-line in Fig.~\ref{fig:1}).

As we can see from Fig.~\ref{fig:1}(b), the cost of connection is dependent on the distribution. The cost is very low for the power law distribution of degrees and power law distribution of intensities correlated to degrees. It is the network topology of scale free networks that arise spontaneously in natural and man made systems \cite{Barabasi2003}. So, the cost of connection is low for the more realistic networks. The cost of connection varies as the topology changes. 
\subsection{Effects of coupling strengths}
\begin{figure}[!ht]
\centering\includegraphics[width=0.99\linewidth]{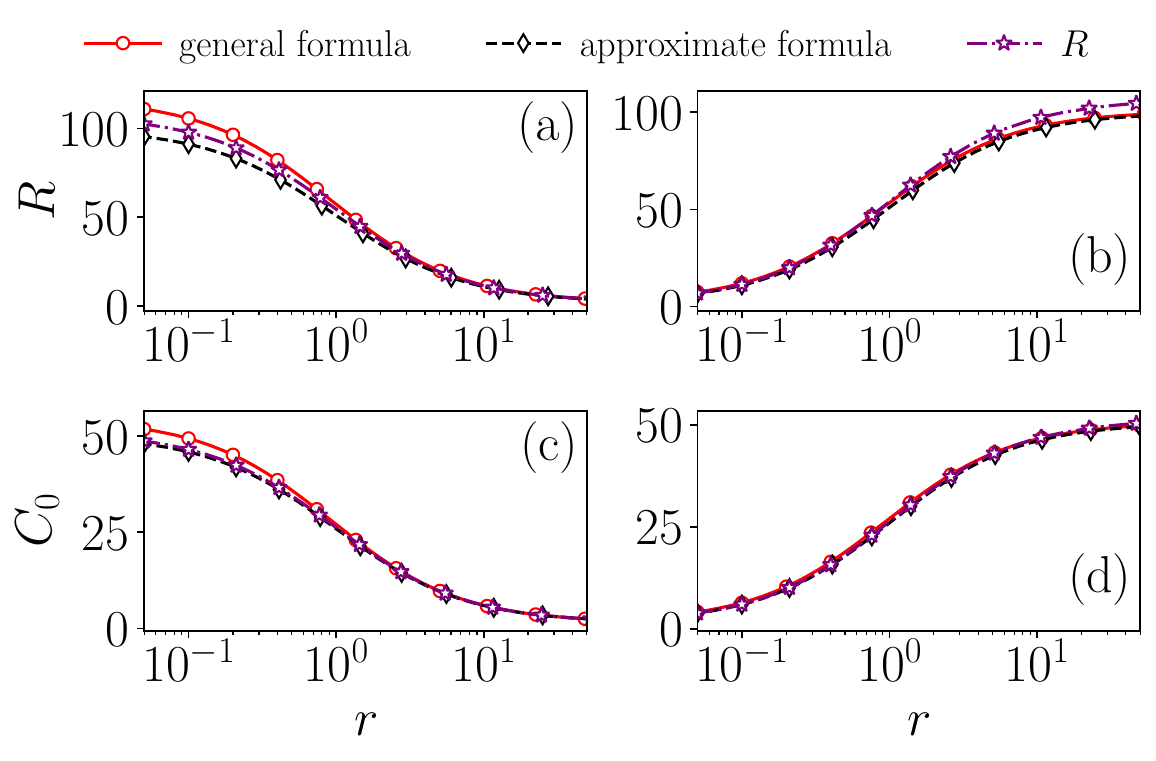} 
\caption{Variation of $R$ as a function of coupling strength ratio $r$ calculated from the general formula, the approximated formula and the eigenvalues of $M$ ($R$), for the simplicial complex when degrees and intensities are uniformly distributed: (a) and (c): $S^{(1)}_{\mbox{max}} > S^{(2)}_{\mbox{max}}$, and (b) and (d): $S^{(2)}_{\mbox{max}} > S^{(1)}_{\mbox{max}}$.}
\label{fig:2}
\end{figure}

Next, we study effect of the coupling strengths of pairwise and three-body interactions. We fix the values of the intensities as $S^{(1)}_{\mbox{min}} = 10$, $S^{(1)}_{\mbox{max}} = 1000$, $S^{(2)}_{\mbox{min}} = 10 $ and $S^{(2)}_{\mbox{max}} = 20$, and and change the values of $r={\sigma_2}/{\sigma_1}$. Here, to change the value of $r$, we fix the pairwise interaction coupling strength $\sigma_1=0.001$ and vary the non-pairwise  coupling strength $\sigma_2$.  It is evident from Fig.~\ref{fig:2}(a) that as the coupling strength for three-body interactions increases, the eigenratio decreases and approaches smaller values and increases the synchronizability of the system. When $r \approx 10$, the general formula gives the exact values for the eigenratio. Then, for larger values of $r$, $R$ from both the approximated and general formulas converge. %
One can easily study the effect of other higher-order interactions using Eq.~\eqref{eq:5.1}. From Fig.~\ref{fig:2}(a), at the small values of $r$, (i.e., $\sigma_2$), the value of the eigenratio approaches to ${S^{(1)}_{\mbox{max}}} / {S^{(1)}_{\mbox{min}}}$. As we increase the $r$, by increasing $\sigma_2$, we can drastically reduce the values of $R$. In Fig.~\ref{fig:2}(a), ${S^{(1)}_{\mbox{max}}}/{S^{(1)}_{\mbox{min}}} = 100 $ and it falls down to $3$ as we increase the three-body interactions. Therefore, higher-order interactions promote synchronization when the first order intensities (degrees in the case of unweighted networks) are very heterogeneous. The same effect is shown in the values of cost as we can see from Fig.~\ref{fig:2}(c). Then we analyze the synchronizability when the intensity of three-body interaction becomes more heterogeneous than the pairwise interaction, i.e., when $S^{(2)}_{\mbox{max}} > S^{(1)}_{\mbox{max}}$. The values are chosen as follows as $S^{(2)}_{\mbox{max}} = 1000$ and $S^{(1)}_{\mbox{max}} = 20 $. In this case, the three-body interaction coupling strength has opposite effect. For a given value of $S^{(2)}_{\mbox{max}} $, the eigenratio increases as we increase the three-body interaction coupling strength $\sigma_2$. Hence, decrease the synchronizability. The values of eigenratio from the approximated formula and the general formula converges when the eigenratio is minimum.

The effect of coupling strength is the same for both cost and the eigenratio [cf. Fig.~\ref{fig:2}(b) and \ref{fig:2}(d)]. The more heterogeneity there is in the second-order intensities, the faster the cost increases. Hence, the synchronizability decreases if we increase the value of $\sigma_2$. The effect of coupling strengths on synchronization depends on the intensity of the network topology. Based on the above results, if $S^{(1)}_{\mbox{max}} > S^{(2)}_{\mbox{max}}$, an increase in $\sigma_2$ enhances synchronizability. However, the opposite scenario occurs if $S^{(2)}_{\mbox{max}} > S^{(1)}_{\mbox{max}}$, where a larger $\sigma_2$ reduces synchronizability. 

In general, higher-order interactions do not always promote synchronization \cite{Zhang2023}. They inhibit the synchronization of the system when the weights of the non-pairwise interactions are higher than those of the pairwise interactions. With this knowledge, we can promote or inhibit synchronization by manipulating the weights and coupling strengths of the pairwise and non-pairwise interactions. 

\subsection{Behaviour of the constants $A_R$, $A_C$, $B_{R_1}$ and $B_{R_2}$}

Next, we analyze the behaviour of the constants $A_R$, $A_C$, $B_{R_1}$, and $B_{R_2}$ as the mean degree changes for a simplicial complex with a mean degree of $K$. For this purpose, we choose the values of the coupling strengths as $\sigma_1 = 0.001$ and $\sigma_2 = 0.01$ and vary ${S^{(1)}_{\mbox{max}}/{S^{(1)}_{\mbox{min}}}} = 1, 2, 10, 100$ by fixing all nodes to the same degree as in the case of $K$-regular networks. Using Eqs. \eqref{eq:15:3} and \eqref{eq:15:4}, we calculate the values of $A_R$, $A_C$, $B_{R_1}$, and $B_{R_2}$ for large values of mean degree $K$. %

\begin{figure}[!ht]
\centering\includegraphics[width=0.99\linewidth]{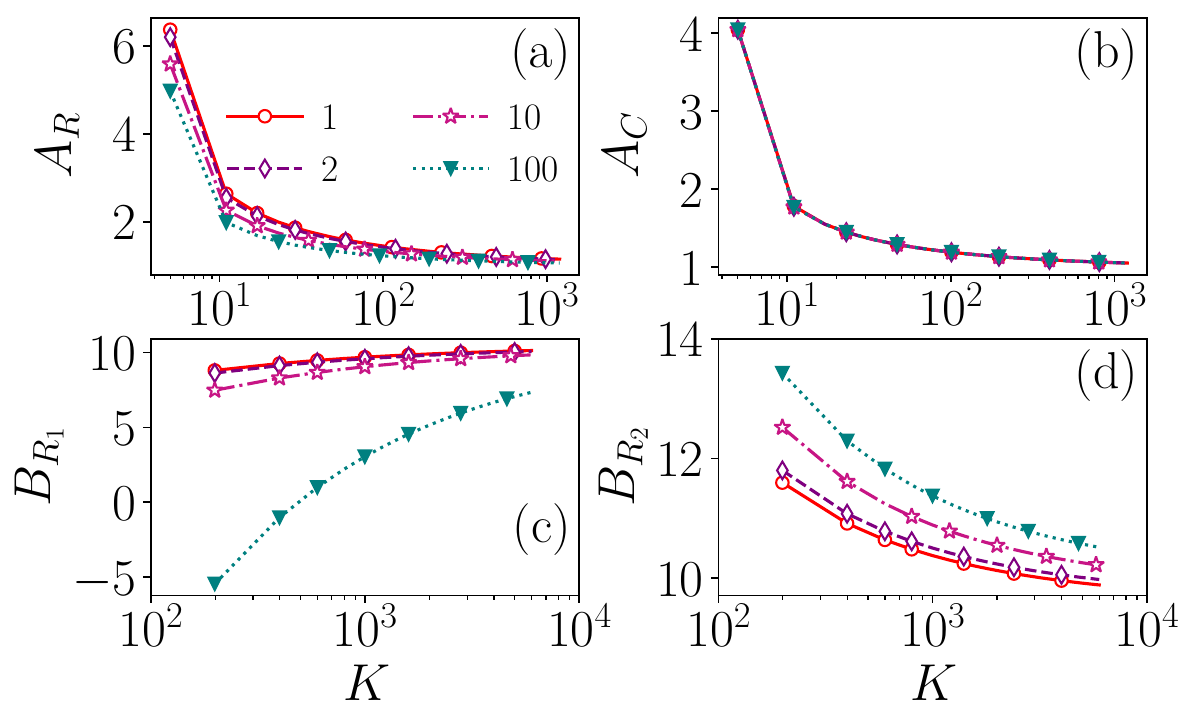} 
\caption{Variations of (a) $A_R$ (b) $A_C$, (c) $B_{R_1}$ and (d) $B_{R_2}$ against $K$. When $K$ is very large, $A_R$ and $A_C$ approach 1 and $B_{R_1}$ and $B_{R_2}$ approach $r$, so we can use the approximate formula rather than the general formula.}
\label{fig:3}
\end{figure}

As we can see from Figs.~\ref{fig:3}(a) and \ref{fig:3}(b), the heterogeneity of the intensities has a small effect on $A_R$, and the behaviour of $A_C$ appears independent of the intensities' heterogeneity as $K$ increases. %
Both $A_R$ and $A_C$ quickly approach unity as $K$ increases.

Further, from Figs.~\ref{fig:3}(c) and \ref{fig:3}(d), we observe that the values of $B_{R_1}$ and $B_{R_2}$ are highly dependent on the heterogeneity of the intensities. $B_{R_1}$ remains close to $r=10$ when the intensities are not very heterogeneous. As the heterogeneity increases, the values decrease for low $K$s and slowly approach $r$. The values of $B_{R_2}$ remain close to the upper bound when the intensities are heterogeneous and slowly approach $r$. When the intensities are homogeneous, i.e., ${S^{(1)}\mbox{max}} / {S^{(1)}\mbox{min}}=1$, the values of $B_{R_2}$ remain close to $r$ for all $K$.

From the behaviours of $A_R$, $A_C$, $B_{R_1}$ and $B_{R_2}$, we can observe that for $K \gg 1$ and not very heterogeneous networks, the general formula approaches the approximated formula. With this, we can easily analyze the synchronizability of the weighted simplicial complexes in the presence of other higher-order interactions, such as four-body interactions, five-body interactions etc., from the knowledge of mean degrees, coupling strengths, and maximum and minimum intensities for the interaction networks.

\subsection{Verification with R\"ossler oscillators}
\label{ross}
To validate our findings, we verify the applicability of the proposed model \eqref{eq:1} or \eqref{eq:intensity}  by analyzing complete synchronization in a system of identical R\"ossler oscillators. In this case, we consider Eq.~\eqref{eq:intensity} to represent the dynamical equations of a weighted simplicial complex composed of R\"ossler oscillators as, %
\begin{subequations}
\label{sys:rossler}
\begin{align}
\dot x_i = & \, -y_i - z_i + \sigma_1\frac{S^{(1)}_i}{K^{(1)}_i}{ \sum_{j=1}^{N}a_{ij}^{(1)}(x_{j} - x_{i})} \notag \\ & + \sigma_2\frac{S^{(2)}_i}{K^{(2)}_i} \sum_{j=1}^N \sum_{k=1}^N a_{ijk}^{(2)}(x_j^2x_k - x_i^3), \\
\dot y_i = & \, x_i + ay_i, \\
\dot z_i = & \, b + z_i(x_i - c), \;\; i = 1,2, \ldots, N, 
\end{align}
\end{subequations}
where $a=0.2$, $b=0.2$ and $c=9.0$, and $N$ is the total number of oscillators. We construct a weighted simplicial complex comprising $50$ R\"ossler oscillators. For a given node $i$, $K^{(1)}_i$ and $K^{(2)}_i$ represent the degrees of pairwise and three-body interactions, respectively, indicating the total number of links and triangles to which the node belongs. $S^{(1)}_i$ and $S^{(2)}_i$ denote the intensities of pairwise and three-body interactions, respectively, for node $i$. The oscillators are coupled through their $x$ components for both pairwise and three-body interactions (later we will show that the results are independent on the choice of components). We fix the values of $\sigma_1$, $\sigma_2$, $S^{(1)}_{\mbox{max}}$, $S^{(2)}_{\mbox{max}}$, $S^{(2)}_{\mbox{min}}$ and vary the values of $S^{(1)}_{\mbox{min}}$. 
\begin{figure*}[!ht]
\centering\includegraphics[width=0.99\linewidth]{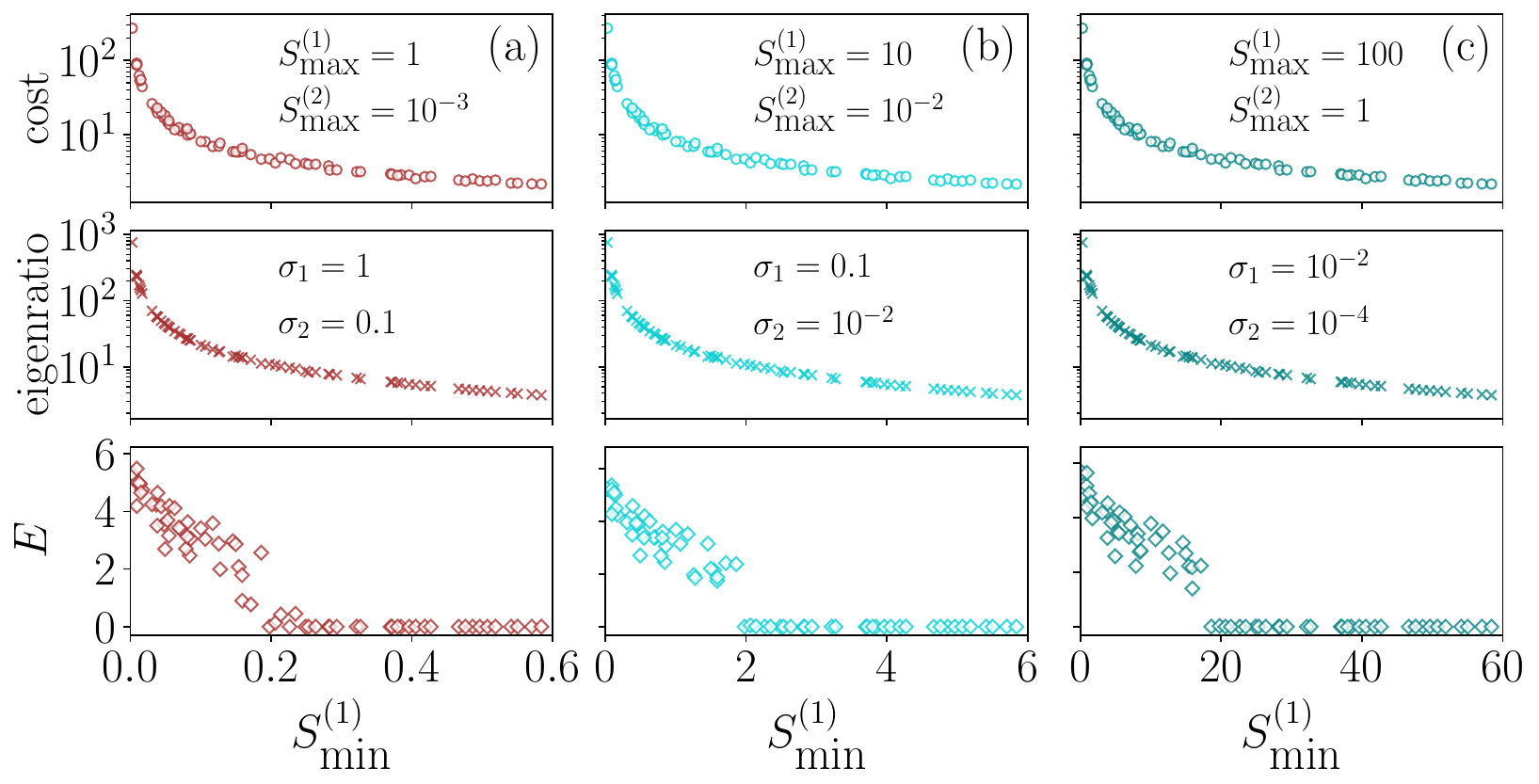} 
\caption{Synchronizability of coupled R\"ossler systems \eqref{sys:rossler} calculated using cost, eigenratio, and synchronization error ($E$) for different values of the coupling strengths, $(S^{(1)}_{\mbox{max}}, S^{(2)}_{\mbox{max}}, \sigma_1, \sigma_2)$: (a) $(1, 10^{-3}, 1,0.1)$ (first column), (b) $(10, 10^{-2}, 0.1,10^{2})$ (second column), and (c) $(100, 1, 10^{-2},10^{-4})$ (third column) chosen as per equation \eqref{eq:29} and $S^{(2)}_{\mbox{min}} = 0.1 \times S^{(2)}_{\mbox{max}}$. The synchronizability of the network is independent of the individual values of the coupling strengths.}
\label{fig:4}
\end{figure*}%
The mean degrees are $K^{(1)} = 28.4$ and $K^{(2)} = 242.1$. To perform the numerical analysis, we randomly distribute the degrees and the intensities. We construct the adjacency tensors as we did previously. We calculate the synchronisation error using the formula \cite{Gambuzza2021},
\begin{align}
E = \left\langle \left(\frac{1}{N(N-1)} \sum_{i,j=1}^N \| \boldsymbol{x}_j - \boldsymbol{x}_i \|^2\right)^{\frac{1}{2}} \right\rangle_T,
\end{align} 
where $\boldsymbol{x}$ is the state of the node, $\langle  \rangle$ represents the time average over time $T$. 

Then, we calculate the eigenratio and cost using the formulas \eqref{eq:15:3} and \eqref{eq:15:4}. We consider three combinations of coupling strengths and maximum intensities, and Fig.~\ref{fig:4} shows the results. We choose the values for the $S^{(1)}_{\mbox{max}}$, $S^{(2)}_{\mbox{max}}$, $\sigma_1$ and $\sigma_2$ as per the relations, 
\begin{subequations} \label{eq:29}
\begin{align} 
S^{(1)}_{\mbox{max}} \times \sigma_1 & \in [1, 2], \label{eq:29a} \\
S^{(2)}_{\mbox{max}} \times \sigma_2 & \in [10^{-4}, 10^0]. \label{eq:29b}
\end{align}
\end{subequations}
We choose the parameters from the range of coupling strengths for synchronization in unweighted simplicial complexes of R\"ossler oscillators \cite{Gambuzza2021} and the results are shown in Fig.~ \ref{fig:4}. We observe that as the eigenratio and cost decrease, the synchronization error also decreases, leading to a more synchronized system. Complete synchronization (i.e., zero synchronization error) occurs when $ S^{(1)}_{\mbox{min}} = 0.2 \times S^{(1)}_{\mbox{max}}$, or 20\% of $ S^{(1)}_{\mbox{max}}$. This indicates that we can assess the synchronizability of complex systems with higher-order interactions using the eigenratio and cost, employing the general formulas \eqref{eq:15:3} and \eqref{eq:15:4} when the system's mean degree is large. Additionally, we observe that a large $S^{(1)}_{\mbox{max}} / S^{(1)}_{\mbox{min}}$ ratio hinders synchronization. Therefore, to inhibit synchronization in systems where its effects are undesirable, we can increase the maximum intensity or decrease the minimum intensity; conversely, to promote synchronization, we can do the opposite.

\begin{figure*}[!ht]
\centering\includegraphics[width=0.99\linewidth]{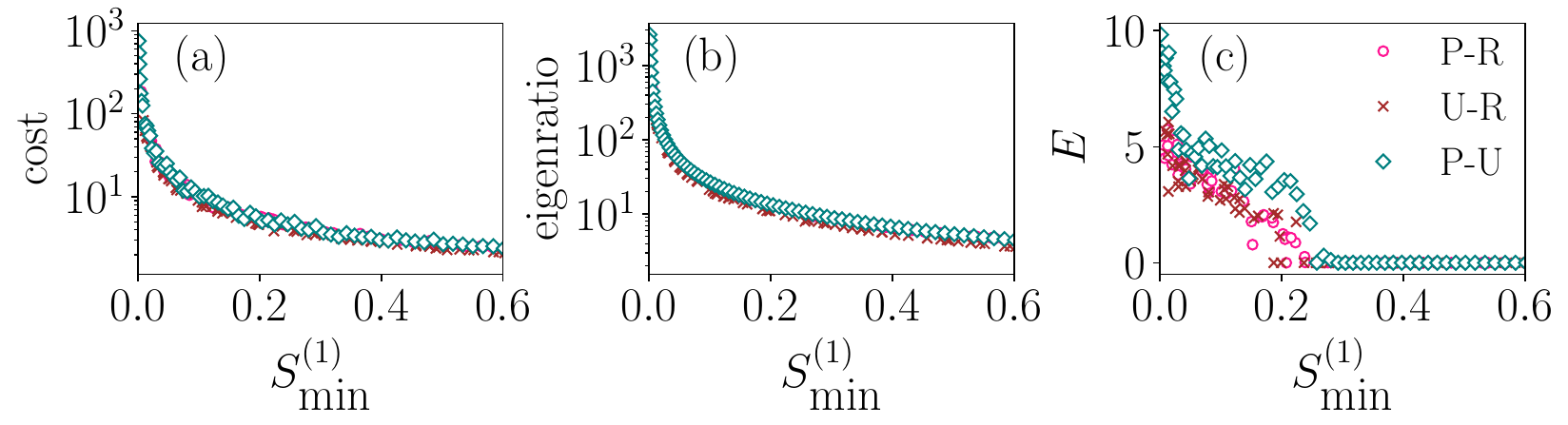} 
\caption{Synchronizability calculated using (a) cost, (b) eigenratio, and (c) synchronization error ($E$) for different distributions of the degrees and the intensities: P-R, U-R, and P-U correspond to the degrees and intensities distributed using power law and random, uniform and random, and power law and uniform, respectively.  The other parameters are $\sigma_1 = 1$,  $\sigma_2 = 0.1$, $S^{(1)}_{\mbox{max}} = 1$, $S^{(2)}_{\mbox{max}} = 10^{-3}$ and $S^{(2)}_{\mbox{min}} = 0.1 \times S^{(2)}_{\mbox{max}}$. System synchronization is independent of the distributions of degrees and intensities.}
\label{fig:5}
\end{figure*}

We validate our formula by conducting additional tests, wherein we select degrees using power-law and uniform distributions. The results are shown in Fig. \ref{fig:5}. We distribute the degrees and intensities as powerlaw and random (P-R), uniform and random (U-R), and powerlaw and uniform (P-U) while maintaining all other parameters as in Fig.~\ref{fig:4}(a). The results consistently demonstrate that synchronization is independent of the connectivity patterns in simplicial complexes. Therefore, the derived formula simplifies the problem by reducing the necessity to compute adjacency tensors for each interaction. Moreover, this formula facilitates a swift and straightforward determination of when a system synchronizes.
\begin{figure*}[!ht]
\centering\includegraphics[width=0.99\linewidth]{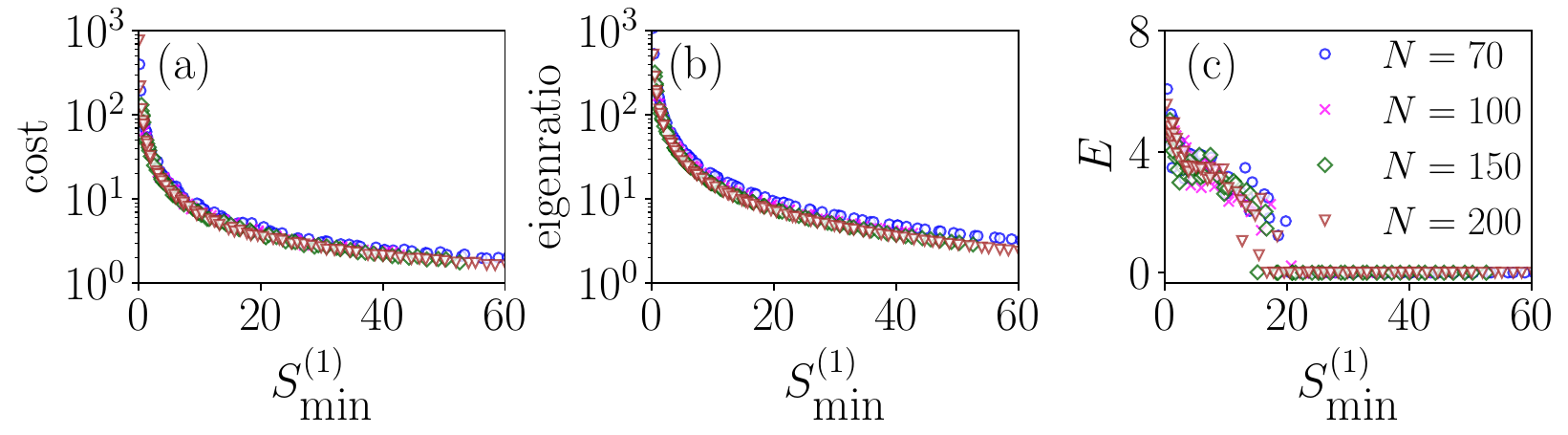} 
\caption{Synchronizability calculated using (a) cost, (b) eigenratio, and (c) synchronization error ($E$) for different system sizes $N = 50, 70, 150, 200$ and $\sigma_1 = 10^{-2} $, $\sigma_2 = 10^{-4}$, $S^{(1)}_{\mbox{max}} = 100$, $S^{(2)}_{\mbox{max}} = 1$, and $S^{(2)}_{\mbox{min}} = 0.1 \times S^{(2)}_{\mbox{max}}$. The complete synchronization of the system is independent of the total number of nodes present in the network.}
\label{fig:6}
\end{figure*}%
We then increase the number of oscillators to $N = 70, 100, 150$ and $200$ while maintaining all other parameters as in Fig.~\ref{fig:4}(b). We observe that synchronization occurs at the same value of $S^{(1)}_{\mbox{min}}$ as in the system with $N = 50$ oscillators. This suggests that complete synchronization in weighted simplicial complexes of identical oscillators is independent of the total number of oscillators in the system. We plot the results in Fig. \ref{fig:6}. Notably, the synchronization errors for  $N = 150$ and $N = 200$ are precisely the same. %
\begin{figure*}[!ht]
\centering\includegraphics[width=0.99\linewidth]{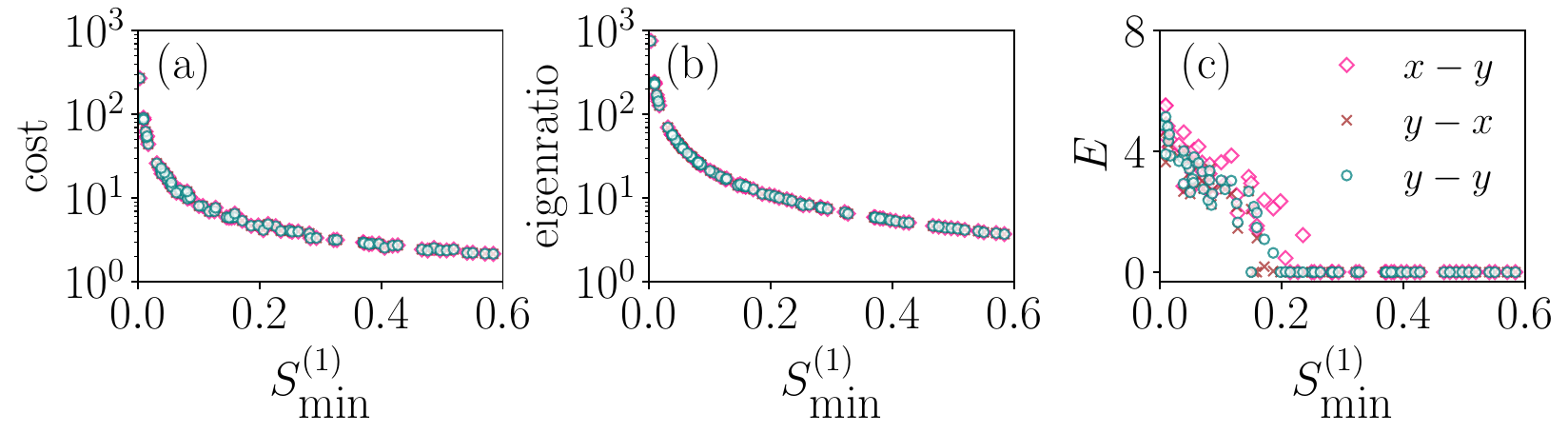} 
\caption{Synchronizability calculated using (a) cost, (b) eigenratio, and (c) synchronization error ($E$) for the interactions happening between different components. Here $\diamond$ represents the result for  $x-y$ scheme (i.e., pairwise interaction through $x$ components and three body interactions through $y$ components), $\times$ for $y-x$ (pairwise interaction through $y$ components and three-body interactions through $x$ variable) and $\circ$ for $y-y$ (both interactions through $y$ components). The other parameters are $\sigma_1 = 1 $,  $\sigma_2 =  0.1$, $S^{(1)}_{\mbox{max}} = 1$, $S^{(2)}_{\mbox{max}} = 10^{-3}$  and $S^{(2)}_{\mbox{min}} = 0.1 \times S^{(2)}_{\mbox{max}}$. The synchronizability of the system is independent of the way the nodes are interacting.}
 \label{fig:7}
\end{figure*}%
Analyzing large weighted complex networks with higher-order interactions poses a major challenge, especially due to the presence of a vast number of nodes in natural and man-made systems. Our general formulas \eqref{eq:15:3} and \eqref{eq:15:4} can significantly simplify the study of complete synchronization in such systems.
 
Next, we check the synchronizability by taking three different coupling functions through different variables. We consider the pairwise interactions through the $x$ components and three-body interactions through the $y$ components ($x-y$ scheme). We then conduct additional experiments with pairwise interactions through $y$ components, three-body interactions through $x$ components ($y-x$ scheme), and both interactions through $y$ components ($y-y$ scheme). All other values remain identical to those used in Fig. \ref{fig:4}(a).  Figure \ref{fig:7} demonstrates that the system synchronizes at the same value of $S^{(1)}_{\mbox{min}}$, indicating that synchronization in weighted simplicial complexes is independent of the specific interaction modes among oscillators.

\subsection{Verification with real-world connectivity structure}\label{ants}
\begin{figure*}[!ht]
\centering\includegraphics[width=0.99\linewidth]{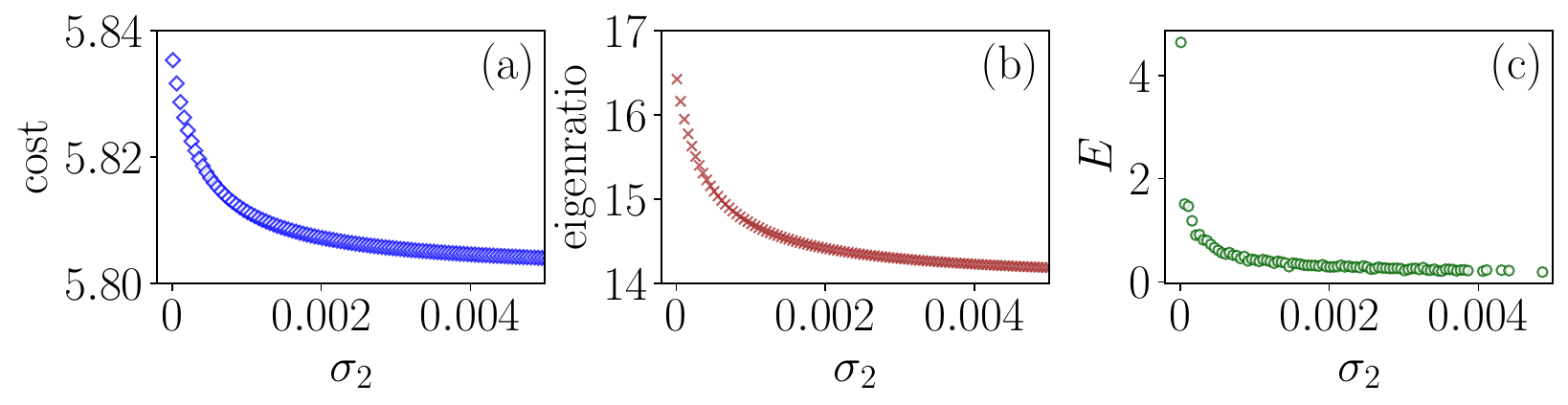} 
\caption{Synchronizability of real-world connectivity network \cite{nr-aaai15} of $N=152$ oscillators modelled by R\"ossler systems calculated using (a) cost, (b) eigenratio, and (c) synchronization error ($E$) by varying the nonpairwise coupling strength $\sigma_2$.}
\label{fig:8}
\end{figure*}

Up to now, we verified our derived analytical theories on synthetic network structures. Finally, we verify our formula on real-world connectivity networks \cite{nr-aaai15}. The system is a collection of ants, where the nodes represent the ants, and the links represent interactions between them within a day. The weight of links represents the frequency of interaction. We consider all triangles as three-body interactions and assign weights to each proportional to the weights of the links in the triangles. \textcolor{blue} {At each time instant, there may be two agents (nodes) that are not connected, i.e., the whole network is disconnected at that time. 
However, the union of all the connections over the entire period of time gives a connected network, which is a necessary condition for complete synchronization in a graph.
The real-world connectivity network} has $N = 152$, $K^{(1)}_{\mbox{max}} = 143$, $K^{(1)}_{\mbox{min}} = 32 $, $K^{(2)}_{\mbox{max}} = 7.1 K$ with mean degree $K^{(1)} = 102 $, average number of triangles $ K^{(2)} = 4.1 K $, $S^{(1)}_{\mbox{max}} = 2064 $, $ S^{(2)}_{\mbox{max}} = 5101.32 $, $ S^{(1)}_{\mbox{min}} = 187 $ and $ S^{(2)}_{\mbox{min}} = 387.62 $. Previously, the oscillatory dynamics was used to represent the dynamics of each node in the modelling of opinion formation in social systems \cite{opinion} since the opinions are not necessarily stationary always. 
In our case, we model the dynamics of the ant interaction network and place chaotic R\"ossler oscillators on that network. 
We chose the values for $\sigma_1$ and $\sigma_2$ according to the relations in Eq.~\eqref{eq:29}. We fix $\sigma_1 = 0.001$ and vary $\sigma_2$ from $10^{-5}$. Figure \ref{fig:8} displays the results.
One may note that as the eigenratio and the cost decrease, the synchronization error also decreases. The system is synchronized when the values of the eigenratio and the cost are at a minimum for the chosen values of the coupling strengths. From these results, we can conclude that it is possible to predict the complete synchronizability of complex systems with higher-order interactions using our formula.

As a whole, it is intriguing to note that complete synchronization in large complex networks is independent of the total number of nodes present in the system and the way they are connected. Synchronization depends on the coupling strengths and the degree of connection between nodes. If all nodes are connected with nearly equal strengths, the system is easily synchronized. However, if some nodes have significantly stronger connections compared to others, achieving synchronization becomes challenging. By knowing the weights of links and triangles, we can calculate intensities. Subsequently, coupling strengths must be chosen based on the relationships mentioned earlier. This allows us to predict when synchronization will occur, which is applicable to a broad range of complex systems.

\section{Conclusion}

In this paper, we have analyzed the nature of synchronization in randomized weighted simplicial complexes of identical oscillators with natural coupling. We derived general formulas for the eigenratio and the cost and verified these theoretical results for various network topologies with diffusive couplings and intensity distributions. Our observations indicated that synchronization is independent of the structure of simplicial complexes, the way the oscillators interact with one another and the total number of oscillators present in the system. We observed complete synchronization for fixed values of $\sigma_1$, $\sigma_2$, $S^{(1)}_{\mbox{max}}$, $S^{(2)}_{\mbox{max}}$, and $S^{(2)}_{\mbox{min}}$, when $S^{(1)}_{\mbox{min}} = 0.2 \times S^{(1)}_{\mbox{max}}$ (i.e., $20$ percentage of $S^{(1)}_{\mbox{max}}$). Notably, the synchronization error values for $N = 150$ and $N = 200$ are nearly identical, indicating that synchronisation is independent of the number of oscillators when $N$ is large. Our general formula provides a good approximation for complete synchronization in these cases. 

Furthermore, our observations highlighted that the effect of coupling strengths varies depending on the largest intensity. When $S^{(1)}_{\mbox{max}} > S^{(2)}_{\mbox{max}}$, an increase in $\sigma_2$ enhances synchronizability. Conversely, when $S^{(2)}_{\mbox{max}} > S^{(1)}_{\mbox{max}}$, synchronization decreases with an increase in $\sigma_2$. The behaviours of constants $A_R$, $A_C$, $B_{R_1}$, and $B_{R_2}$ suggest that the approximated formula, containing only coupling strengths and intensities, can be used.

We have validated our results by applying the general formula to determine the synchronizability of weighted simplicial complexes of diffusively coupled identical R\"ossler oscillators in Sec. \ref{ross}. We further verified our results using real-world connectivity structure where the R\"ossler oscillators are placed on each node in Sec. \ref{ants}. From numerical simulations in Sec. \ref{ross} and \ref{ants}, we see that the results are robust on the number of nodes, components through which they are coupled and the distributions of degree and intensities.

In conclusion, we demonstrated that the synchronizability of large randomized weighted simplicial complexes with natural coupling can be determined using mean degrees, coupling strengths, and intensities without the need to construct connectivity matrices or explicitly calculate their eigenvalues. These formulas substantially reduce computation time and cost when assessing a system's tendency to synchronize. In essence, these formulas enable the prediction and control of synchronizability in complex networks with higher-order interactions, achieved by manipulating the degrees, weights, and coupling strengths.

\acknowledgements
The work of S.N.J. and P.M. is supported by MoE RUSA 2.0 (Bharathidasan University - Physical Sciences).


%

\end{document}